\documentclass[fleqn,usenatbib]{mnras}
\usepackage{newtxtext,newtxmath}
\usepackage{mathptmx}
\usepackage{txfonts}

\usepackage[T1]{fontenc}
\usepackage{aecompl}
	
\usepackage{graphicx}	
\usepackage{amsmath}	
\usepackage{amssymb}	
\usepackage{tabularx}
\usepackage{subfigure}

\def\ha{$H\alpha$}

\def\hb{$H\beta$}

\def\oiii{{[O\sc{iii}]}\/}
\def\oiiia{{[O\sc{iii}]}$\lambda$4959\/}
\def\oiiib{{[O\sc{iii}]}$\lambda$5007\/}
\def\oiiill{{[O\sc{iii}]}$\lambda\lambda$4959,5007\/}

\def\niia{{[N\sc{ii}]}$\lambda$6548\/}
\def\niib{{[N\sc{ii}]}$\lambda$6583\/}
\def\sii{{[S\sc{ii}]}$\lambda\lambda$6716,6731\/}
\def\siia{{[S\sc{ii}]}$\lambda$6716\/}
\def\siib{{[S\sc{ii}]}$\lambda$6731\/}

\def\oia{{[O\sc{i}]}$\lambda$6300\/}

\def\l{$\lambda$}

\def\l5100{$L_{\it 5100}$}

\title[Possible binary AGN in Mrk\,622]
{A possible binary AGN in Mrk\,622?} 

\author[E. Ben\'{\i}tez et al.]
{E.~Ben\'{\i}tez,$^{1}$\thanks{Email: erika@astro.unam.mx}
J.~M.~Rodr\'iguez-Espinosa,$^{2,3}$
I.~Cruz-Gonz\'alez,$^{1}$
O.~Gonz\'alez-Mart\'in,$^{4}$
\newauthor
C.~A.~Negrete,$^{5}$
D.~Ruschel-Dutra,$^{6}$
L.~Guti\'errez,$^{7}$ and 
E.~Jim\'enez-Bail\'on $^{7}$
\\
$^{1}$Instituto de Astronom\'{\i}a, Universidad Nacional Aut\'onoma de M\'exico, AP 70-264, CDMX 04510, Mexico\\
$^{2}$Instituto de Astrof\'isica de Canarias (IAC), V\'ia L\'actea, s/n, 38205, La Laguna, Spain\\
$^{3}$Departamento de Astrof\'isica, Universidad de La Laguna (ULL), 38205, Spain\\
$^{4}$Instituto de Radioastronom\'ia y Astrof\'isica (IRyA-UNAM), 3-72 (Xangari), 8701, Morelia, Mexico\\
$^{5}$CONACYT Research Fellow - Instituto de Astronom\'{\i}a, Universidad Nacional Aut\'onoma de M\'exico, AP 70-264, CDMX 04510, Mexico\\
$^{6}$Departamento de F\'isica, Universidade Federal de Santa Catarina, P.O. Box 476, 88040-900, Florian\'opolis, SC, Brazil \\
$^{7}$Instituto de Astronom\'ia, Universidad Nacional Aut\'onoma de M\'exico, Apdo. Postal 877, Ensenada, 22800 Baja California, Mexico}

\date{Accepted XXX. Received YYY; in original form ZZZ}

\pubyear{2017}
\begin{document}
\label{firstpage}
\pagerange{\pageref{firstpage}\pageref{lastpage}}
\maketitle

\begin{abstract}

Mrk\,622 is a Compton Thick AGN and a double-peaked narrow emission line galaxy, thus a dual AGN candidate. In this work, new optical long-slit spectroscopic observations clearly show that this object is rather a triple peaked narrow emission line galaxy, with both blue and red shifted narrow emission lines, as well as a much narrower emission line centred at the host galaxy systemic velocity. The average velocity offset between the blue and red shifted components is $\sim$500 km\,s$^{-1}$, which is producing the apparent double-peaked emission lines. These two components are in the loci of AGN in the Baldwin, Phillips \& Terlevich (BPT) diagrams and are found to be spatially separated by $\sim$76 pc. Analysis of the optical spatially resolved spectroscopic observations presented in this work favours that Mrk\,622 is a system consisting of a Composite AGN amidst a binary AGN candidate, likely the result of a recent merger. This notwithstanding, outflows from a starburst, or single AGN could also explain the triple nature of  the emission lines.

\end{abstract}


\begin{keywords}
\textcolor{red}{galaxies: active-- quasars:individual: Mrk\,622 -- quasars: emission lines -- galaxies: Seyfert, galaxies: binary AGN  
}
\end{keywords}



\section{Introduction}\label{intro}

In the current $\Lambda$CDM paradigm galaxies grow hierarchically through mergers. During these events, smaller galaxies build more massive galaxies. Observations show that super massive black holes (SMBHs) are common in bulge-dominated galaxies. So, hierarchical structure formation implies that some galaxies should harbour two (or more) SMBHs of mass $\sim$10$^{6}-10^{9}$M$_\odot$ in their centre as the result of a recent merger \citep[e.g.][]{1980Natur.287..307B, 2001ApJ...563...34M,2002MNRAS.331..935Y,2003ApJ...582..559V}. 
Merger-remnant galaxies with two SMBHs in their centres, known as dual AGN \citep[see][]{2007ApJ...660L..23G} should therefore be widespread. Mergers are very efficient at funnelling gas for both star formation and accretion onto SMBHs, according to numerical simulations \citep[e.g.][]{2005ApJ...620L..79S}. This is a relevant process since understanding the onset of Active Galactic Nuclei (AGN)  and how star formation is quenched constitute key ingredients to understand how galaxies evolve. 
Observationally, spectroscopic surveys like the Sloan Digital Sky Survey (SDSS) produced hundreds of candidates of nearby (z$<$0.1) dual AGN, 
\citep[e.g.][]{2009ApJ...705L..76W,2012ApJS..201...31G} 
based in the idea that the presence of double-peaked narrow emission lines was an indication of objects harbouring two SMBHs. Although some of these objects turn out to be true dual AGN based on X-ray \citep{2012ApJ...746L..22K,2016ApJ...824L...4K} 
and radio \citep{2014Natur.511...57D,2016ApJ...826..106G} observations, double-peaked narrow emission lines can also be the result of bulk motion of ionised gas clouds in the form of biconical outflows \citep[e.g.][]{2011ApJ...727...71F} or winds, since in AGN they operate on spatial scales coincident with circumnuclear star formation \citep[see][]{2015ApJ...799...83C}. Also they could be due to rotating gaseous disks \citep{2012ApJ...752...63S} or jet-driven outflows \citep{2010ApJ...716..131R}. Therefore, a direct association between objects with double-peaked narrow emission lines with dual or binary (i.e., SMBHs have $\sim$\,kpc or $\sim$\,pc separations, respectively) AGN is not straightforward. As a result, the number of bona fide dual and/or binary AGN is still rather small, $\sim$20 spatially and spectrally confirmed duals \citep[see][]{2015ApJ...811...14M}. AGN that have been previously observed in mergers with double nuclei 
\citep[e.g.][]{2017arXiv170508556B,2016ApJ...824L...4K} 
can be found with their engines turned on before coalescence since both  AGN are active when the separation of the nuclei is below $\sim$1-10\,kpc \citep[see][]{2012ApJ...748L...7V}. However, much of the AGN activity in mergers do not happens simultaneously and there have been found systems where only one AGN is turned on. 

Mrk\,622, also known as UGC\,4229, is at z=0.023 and is classified as a S0 pec. Seyfert 2 (Sy2) galaxy according to NED.\footnote{The NASA/IPAC Extragalactic database (NED) is operated by the Jet Propulsion Laboratory, California Institute of Technology, under contract with the National Aeronautics and Space Administration.} Among the first spectroscopic studies of this AGN, stands out the work done by \citet{1981ApJ...250...55S}, who found that this galaxy shows emission lines with multiple components. These authors noticed that the profiles of the \oiiill \AA\ have a nearly rectangular shape and Full Width at Half Maximum FWHM$\,=\,$1050$\pm$150 km\,s$^{-1}$, while the profiles of H$\alpha$ and [\ion{N}{ii}]$\lambda\lambda$6548,6583\,\AA\ were more Gaussian with a FWHM$\,=\,$350$\pm$75 km\,$s^{-1}$. In a recent study done with \textit{Herschel} PACs \citep[see][]{2012ApJ...755..171S}, it is found that Mrk\,622 is a composite AGN, i.e. AGN+SB. The \textit{Spitzer}/IRS spectrum shows that it has a Sy2 mid-IR spectrum \citep{2007ApJ...671..124D} with strong PAH emission features. This source has also been observed with \emph{XMM}-Newton and recently proposed to be a Compton-thick (CT) source \citep[see][]{2005A&A...444..119G,2014A&A...569A..71C}. 

In this paper new high signal to noise optical spectroscopy obtained with the Intermediate dispersion Spectrograph and Imaging System \textit{ISIS}, attached to the William Herschel Telescope \textit{WHT} is presented. The aim of these observations is to check for the presence of a double-nuclei in its centre, i.e. to confirm whether this object is a dual or a binary AGN candidate or, on the contrary, if the double-peaked lines are due to the kinematics associated to the narrow-line region in the form of winds/outflows or disk rotation. The new optical data presented in this work allowed us to establish that the object shows triple-peaked narrow emission lines, and that this feature is also confirmed after re-modelling a previous SDSS-DR7 spectrum. 

The  paper is organised as follows: In Sect.~\ref{optical} the \textit{WHT} optical spectroscopic observations and the reduction process are described. In Sect~\ref{spectral} the spectral analysis and modelling of the optical \textit{WHT} and \textit{SDSS} data is presented. The discussion and conclusions are finally given in Sect.~\ref{dis}. The  cosmology adopted  in  this work  is $H_{0}$\,=\,69.6~km/,s$^{-1}$\,Mpc$^{-1}$, $\Omega_{m}$\,=\,0.286 and $\Omega_{\lambda}$\,=\,0.714 \citep{2014ApJ...794..135B}.

\section{Optical spectroscopic observations and data reduction}
\label{optical}

Optical long-slit data were obtained using the Intermediate
dispersion Spectrograph and Imaging System (ISIS), attached to the 4.2-m William Herschel Telescope (\textit{WHT}) at the Roque de los Muchachos Observatory. ISIS has two CCD arrays, an EEV12 for the blue arm and a  REDPLUS CCD for the red arm. The blue CCD was centred around 4500 \AA, and the red one at 6999 \AA. The gratings used were R600B and R600R which provide a dispersion of 0.44 and 0.49 \AA\,pixel$^{-1}$, respectively. {The slit width was set to 1.018\,\arcsec, this is about 3.3 pixels FWHM.  The spectral resolution in the blue region is 2.06 \AA\, ($\sim$\,123\,km\,s$^{-1}$) and in the red 1.84 \AA\, ($\sim$\,84\,km\,s$^{-1}$). The spatial sampling along the slit is 0.20 and 0.22\,\arcsec\,pixel$^{-1}$ in the blue and red spectral regions, respectively. 
The slit was placed at a position angle $\sim$-92$^{\circ}$ and observations were done with seeing ranging between 1\arcsec and 1.5\arcsec. With this set up, two exposures of 
1200\,s in the blue (3550-5250 \AA), and in the red (5860-7780 \AA) were obtained. Data were reduced and calibrated using standard \textit{IRAF} packages\footnote{IRAF is distributed  by the  National  Optical Astronomy  Observatories, operated by the Association of Universities for Research in Astronomy, Inc., under cooperative  agreement with  the  National Science Foundation}. Bias subtraction, flat-fielding, wavelength and flux calibration were done to both spectra. For the wavelength calibration, CuAr and HeNeAr lamps were used. The standard star Feige\,34 from the \citet{1990AJ.....99.1621O} catalogue was used for flux calibration. Sky subtraction was done using task \textit{IRAF/Background}. Spectra were combined to produce a final 2400\,s exposure time spectrum with a mean airmass of 1.22. An aperture of 3\arcsec\, was used to extract the 1-dimensional spectrum. Table \ref{obs} shows the log of the \textit{WHT} observations as well as of archival \textit{SDSS-DR7} data obtained for Mrk\,622 that will be re-analysed in this work.

\begin{table}
\caption{Log of observations}
\label{obs}
\begin{tabular}{lccc}
\hline 
\hline 
Telescope  & \textit{WHT}  & \textit{SDSS}-DR7 \\
\hline
Observation Date	&	01-28-2015	&	12-11-2001	\\
Instrument  & ISIS & SDSS Spectrograph\\
S/N(5100 \AA) & 29 & 41  \\	
S/N(6425 \AA) & 55 & 84 \\
Exposure time (s) & 2400  & 2700 \\
\hline									
\hline \\
\end{tabular}
\end{table}

\section{Spectral Analysis}
\label{spectral}
The Starlight code \citep{2005MNRAS.358..363C,2006MNRAS.370..721M} was used to model the contribution of the host galaxy using a combination of simple stellar population models from \citet{2003MNRAS.344.1000B}. The spectrum of Mrk\,622 was corrected for Galactic extinction using the dust maps by \citet{1998ApJ...500..525S} and the extinction law of \citet{1989ApJ...345..245C}. The most prominent emission lines were masked following the same procedure described in \citet{2013ApJ...763...36B}. Since the AGN continuum contribution resulted to be negligible, only the resultant stellar continuum fit was subtracted to the \textit{WHT} spectrum, in order to produce a pure AGN emission line spectrum. The spectral decomposition is shown in the upper panel of Figure~\ref{host}.  

The systemic velocity for the host galaxy was derived from the Balmer absorption lines by means of the penalized pixel fitting method  \citep[pPXF,][]{2004PASP..116..138C}. This technique fits the host galaxy spectrum with a synthetic stellar population convolved with a kinematic model, which is defined by a Gauss-Hermites series up to the fourth order \citep{1993ApJ...407..525V}. The stellar population model consists of a linear combination of simple stellar populations from the \citet{2010MNRAS.404.1639V} library. To estimate the kinematic parameters a portion of the spectrum between 3810\,\AA\ and 4250\,\AA\ was used (see lower panel of Figure~\ref{host}). Additionally, a smooth polynomial function of third order was used to remove low frequency oscillations in the continuum level. The uncertainties in the best fit parameters were derived from 1000 realizations of the fit, each time having the flux randomized by a normal distribution with a standard deviation equivalent to a signal to noise ratio of 50. The fits to the spectrum result in a systemic velocity of 6989.1\,$\pm$\,2.7 km\,s$^{-1}$, with a velocity dispersion $\sigma$ of 178\,$\pm$\,9 km\,s$^{-1}$.  From our estimation of the bulge stellar velocity dispersion and using the black hole mass M$_{BH}$ vs. $\sigma$ relation given by \citet{2013ApJ...764..184M}, an upper limit value to the total BH mass of the black holes yields  $\log$ M$_{BH}$\,=\,7.50\,$\pm$ 0.12\,M$_\odot$.

\begin{figure}
\subfigure{\includegraphics[width=8.5cm,height=5cm]{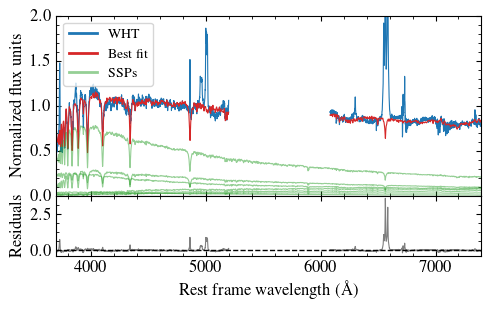}}
\subfigure{\includegraphics[width=8.5cm,height=5cm]{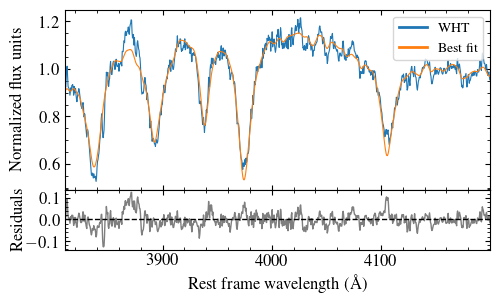}} \\
\caption{Upper panel: stellar population decomposition done to the combined \textit{WHT} spectrum using starlight. At the bottom, the pure emission line spectrum obtained after removing the best fit is shown. Lower panel: pPXF fit to the portion of the spectrum used for determining the systemic velocity of the host galaxy, at the bottom the residuals are shown.}
\label{host}
\end{figure}

Since Mrk\,622 has been previously found to be a double-peaked emission line object \citep[see][]{2009ApJ...705L..76W}, the profile fitting was done initially assuming two Gaussian components or 2G model. The spectral analysis was done using \textit{IRAF/specfit} \citep{1994ASPC...61..437K}. With this code the spectrum was fitted in three regions: region 1, from 4820 to 5040 \AA, region 2 from 6500 to 6770\,\AA, and region 3 from 6280 to 6380\,\AA. The \oiiill \AA\ lines were fitted first. Assuming that all Narrow Line Region (NLR) lines have the same physical origin, the same FWHM in the 2G model was used to fit the \oiiill \AA\ lines. The same criteria was applied to the rest of the narrow lines in the blue and red spectral regions. The reduced $\chi^2$ values for regions 1, 2 and 3 obtained with \textit{Specfit} and the 2G model are: 1.534, 0.191 and 0.518, respectively. Although the best fit was obtained with the 2G model, a broad component with a FWHM of $\sim$\,1700\,km\,s$^{-1}$ was necessary to fit H$\alpha$. The 2G model resulted very contrived considering that Mrk\,622 is a Sy2 galaxy, thus the 2G model was discarded. To check this preliminary result, the 2-dimensional (2D) spectra obtained with the \textit{WHT} were revisited. The new analysis shows that there are three clearly spatially separated emitting regions in \oiiill \AA\,(see Figure~\ref{triple}). These regions can be seen in each of the two spectra obtained before combining them. Between the blue and red shifted components the angular projected spatial offset is $\sim$0.81$\pm$0.4 pixels or $\sim$0.16\,\arcsec, {\it i.e.} $\sim$76.5\,pc. The spatial offset was estimated with \textit{IRAF/IMCENTROID}. Based on this result, we decided to perform a new fit but this time using  three Gaussian components or a 3G model. The reduced $\chi^2$ values for regions 1, 2 and 3 obtained with the 3G model are: 1.334, 0.158 and 0.548, respectively. The results thus obtained are shown in the upper panels of Figure~\ref{3Gwht}.  All fits were done via $\chi^{2}$ minimisation using the corresponding algorithm from \textit{Numrecipes}. With the 3G model, a central Gaussian component is always found to be at the restframe, defined by the systemic velocity. In the blue region, the central Gaussian component resulted to be the narrowest of the three, having a FWHM of 234\,$\pm$\,16\,km\,s$^{-1}$. The other two blue and red shifted components have FWHM\,=\,601\,$\pm$\,40\,km\,s$^{-1}$ and FWHM\,=\,427\,$\pm$\,19\,km\,s$^{-1}$. In the red region, the central Gaussian component has a FWHM of 210\,$\pm$\,10\,km\,s$^{-1}$, and a FWHM of 528\,$\pm$\,64\,km\,s$^{-1}$ and 603\,$\pm$\,162\,km\,s$^{-1}$ for the blue and red shifted components, respectively. These components are shown in blue and red colours in the upper panels of Figure~\ref{3Gwht}. The velocity offsets $\Delta {V}$ between the blue and red shifted components were estimated with respect to the restframe component and presented in Table~\ref{off}. Based on the results discussed above, the 3G model is the best model for fitting the three peaks in Mrk\,622, see Table~\ref{off}.

\begin{figure}
\includegraphics[width=8.5cm]{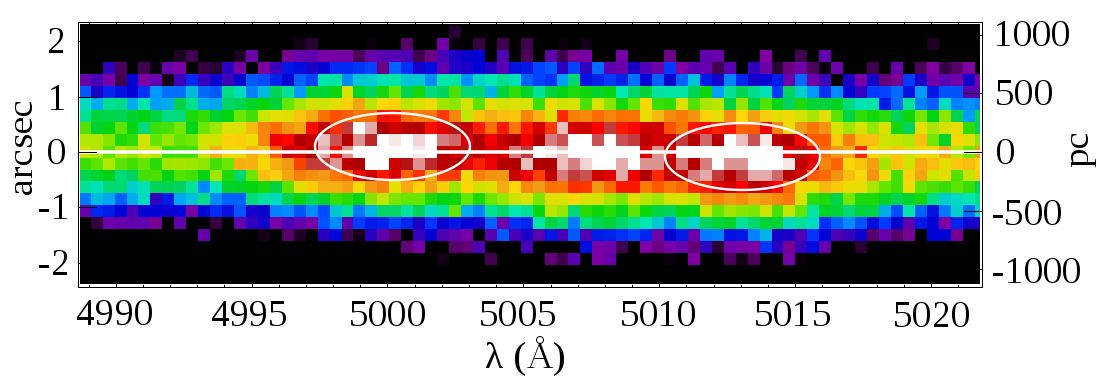}
\caption{A blow-up of the 2D long-slit spectrum obtained with the \textit{WHT}. The \oiiib \AA\ emission line region is shown. The three peaks associated to this emission line are clearly seen, as well as a small vertical shift in the centroids of the two blue and red emission lines. The spatial scale is 0.20\,\arcsec pixel$^{-1}$. The centroid positional shift of the blue and red peaks (marked with two ellipses) is 0.81 pixel ($\sim$\,162 milliarcsec), corresponding to $\sim$76 pc at the distance of Mrk\,622.}
\label{triple}
\end{figure}

\begin{table}
\caption{Velocity offsets $\Delta {V}^{a}$}
\label{off}
\begin{tabular}{lcccc}
\hline
\hline 
Line  & $\Delta{V}^{a}$(\textit{WHT}) & $\Delta{V}^{a}$(\textit{SDSS})\\
&  (km\,s$^{-1}$) & (km\,s$^{-1}$) \\
\hline
\hb	    &	449\,$\pm$\,140	&	485\,$\pm$\,142\\
\oiiia	&	707\,$\pm$\,~31	    &	721\,$\pm$\,~50\\
\oiiib	&	707\,$\pm$\,~31	    &	721\,$\pm$\,~50\\
\ha	    &	430\,$\pm$\,~83	    &	428\,$\pm$\,~39\\
\niia	&	387\,$\pm$\,148	&	272\,$\pm$\,~52\\
\niib	&	444\,$\pm$\,135	&	451\,$\pm$\,~42\\
\siia	&	422\,$\pm$\,122	&	399\,$\pm$\,~90\\
\siib	&	523\,$\pm$\,127	&	365\,$\pm$\,123\\
\oia	&	567\,$\pm$\,~97	    &	358\,$\pm$\,153\\
Average	&	515\,$\pm$\,102	&	467\,$\pm$\,~82	\\
\hline
\multicolumn{3}{l}{$^{a}$ $\Delta{\rm V}=\Delta{V(B)}-\Delta{V(R)}$. B=blue, R=red.} \\
\end{tabular}
\end{table}

\begin{table}
\caption{Intensity ratios with 3G model}
\label{ratios}
\setlength{\tabcolsep}{2pt}
\begin{tabular}{lcccccc}
\hline 
\hline 
Intensity ratio$^{a}$ & \textit{WHT} & \textit{SDSS}\\
\hline 
$\log$(\oiiib/\hb)C	&  -0.19\,$\pm$\,0.15 &  -0.33\,$\pm$\,0.22 \\
$\log$(\oiiib/\hb)B	&	0.76\,$\pm$\,0.12 &   0.75\,$\pm$\,0.11 \\
$\log$(\oiiib/\hb)R	&	0.92\,$\pm$\,0.18 &   0.72\,$\pm$\,0.24 \\
\hline 
$\log$(\niib/\ha)C	&  -0.10\,$\pm$\,0.03 &   -0.08\,$\pm$\,0.05 \\
$\log$(\niib/\ha)B	&	0.10\,$\pm$\,0.12 &    0.08\,$\pm$\,0.10 \\
$\log$(\niib/\ha)R	&  -0.03\,$\pm$\,0.17 &   -0.07\,$\pm$\,0.11 \\	
\hline 
$\log$(\sii/\ha)C	&	-0.48\,$\pm$\,0.06 &  -0.36\,$\pm$\,0.09  \\
$\log$(\sii/\ha)B	&	-0.48\,$\pm$\,0.29 &  -0.39\,$\pm$\,0.14  \\	
$\log$(\sii/\ha)R	&	-0.42\,$\pm$\,0.23 &  -0.45\,$\pm$\,0.26  \\
\hline 
$\log$(\oia/\ha)C	&	-1.22\,$\pm$\,0.06 &  -1.26\,$\pm$\,0.11  \\
$\log$(\oia/\ha)B	&	-0.89\,$\pm$\,0.11 &  -0.98\,$\pm$\,0.22  \\
$\log$(\oia/\ha)R	&	-1.17\,$\pm$\,0.08 &  -0.96\,$\pm$\,0.21  \\
\hline 
\multicolumn{2}{l}{$^{a}$ Central(C), blue (B) and red (R) components.} \\
\end{tabular}
\end{table}

\begin{figure*}
\centering
\subfigure{\includegraphics[width=5.3cm,height=5cm]{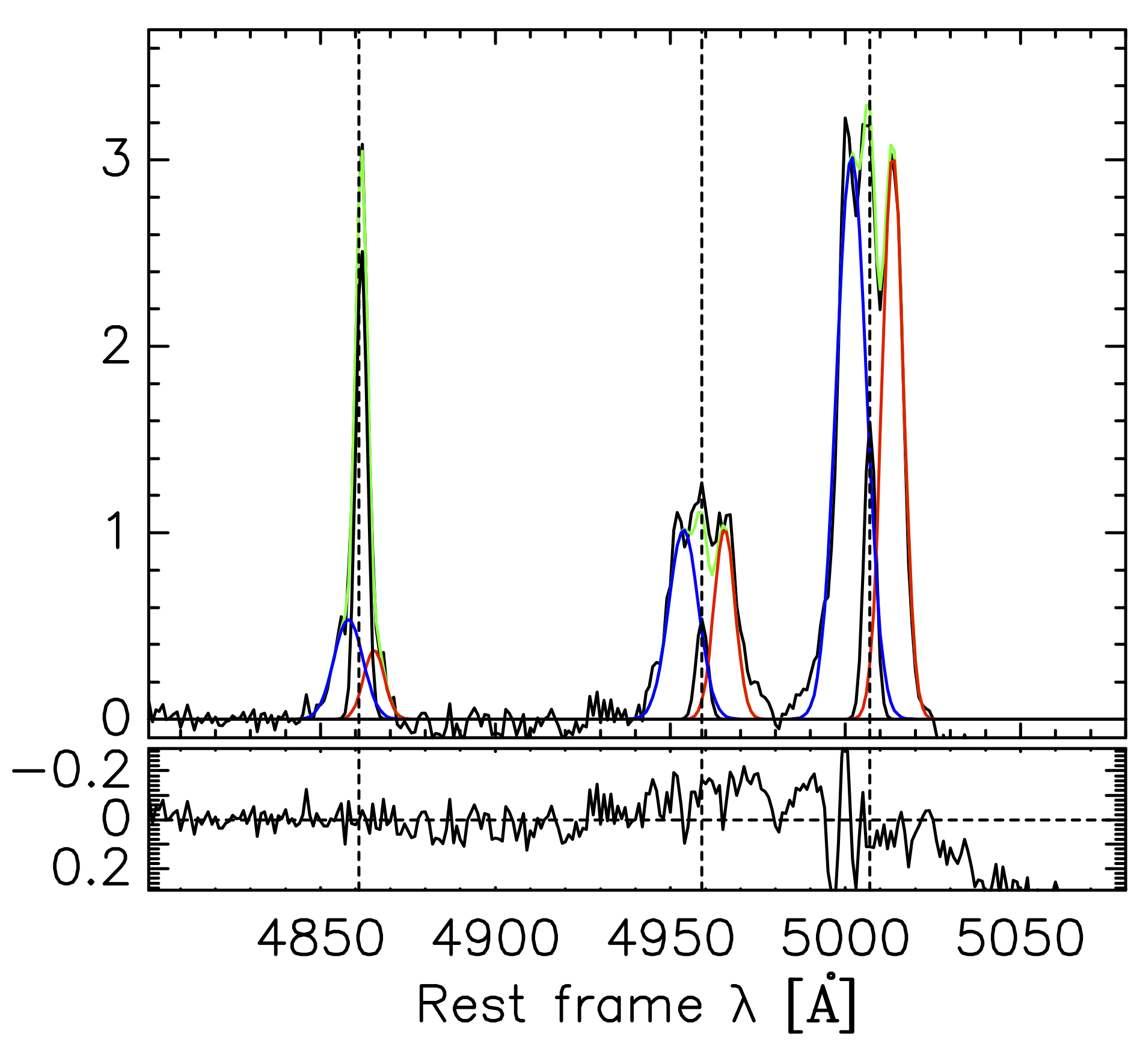}} \subfigure{\includegraphics[width=5.3cm,height=5cm]{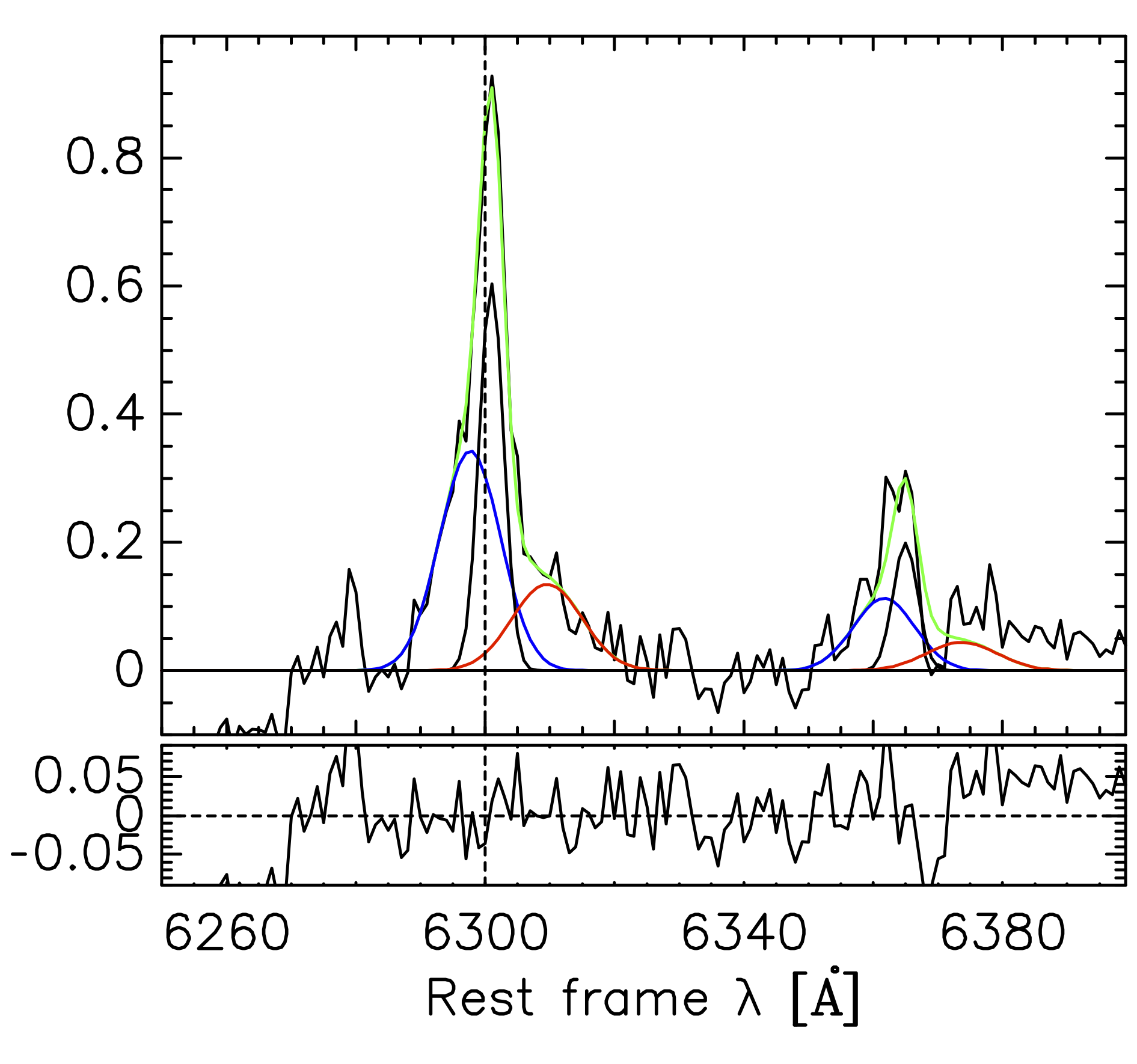}} \subfigure{\includegraphics[width=5.3cm,height=5cm]{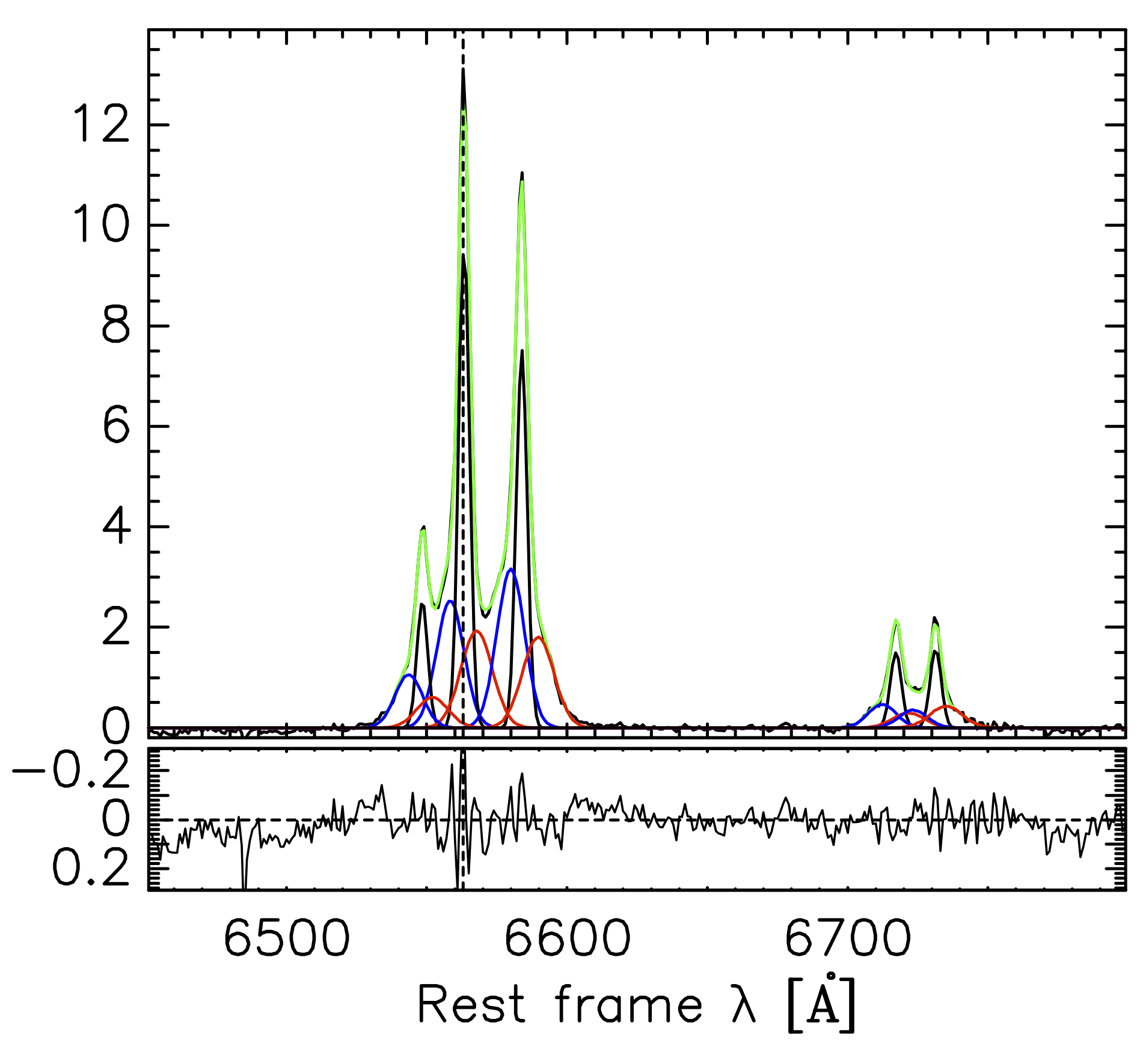}}\\
\subfigure{\includegraphics[width=5.5cm,height=5cm]{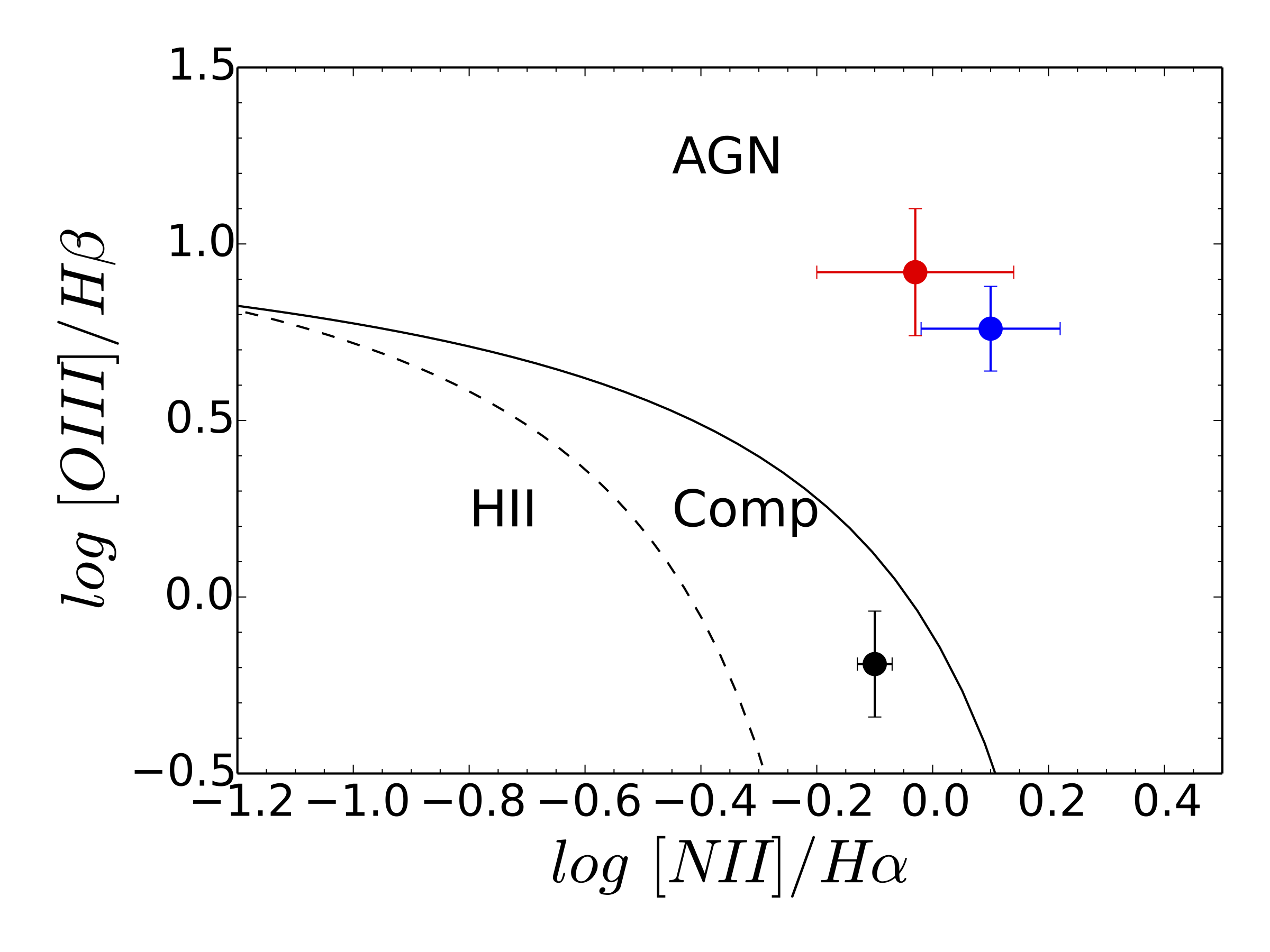}} 
\subfigure{\includegraphics[width=5.5cm,height=5cm]{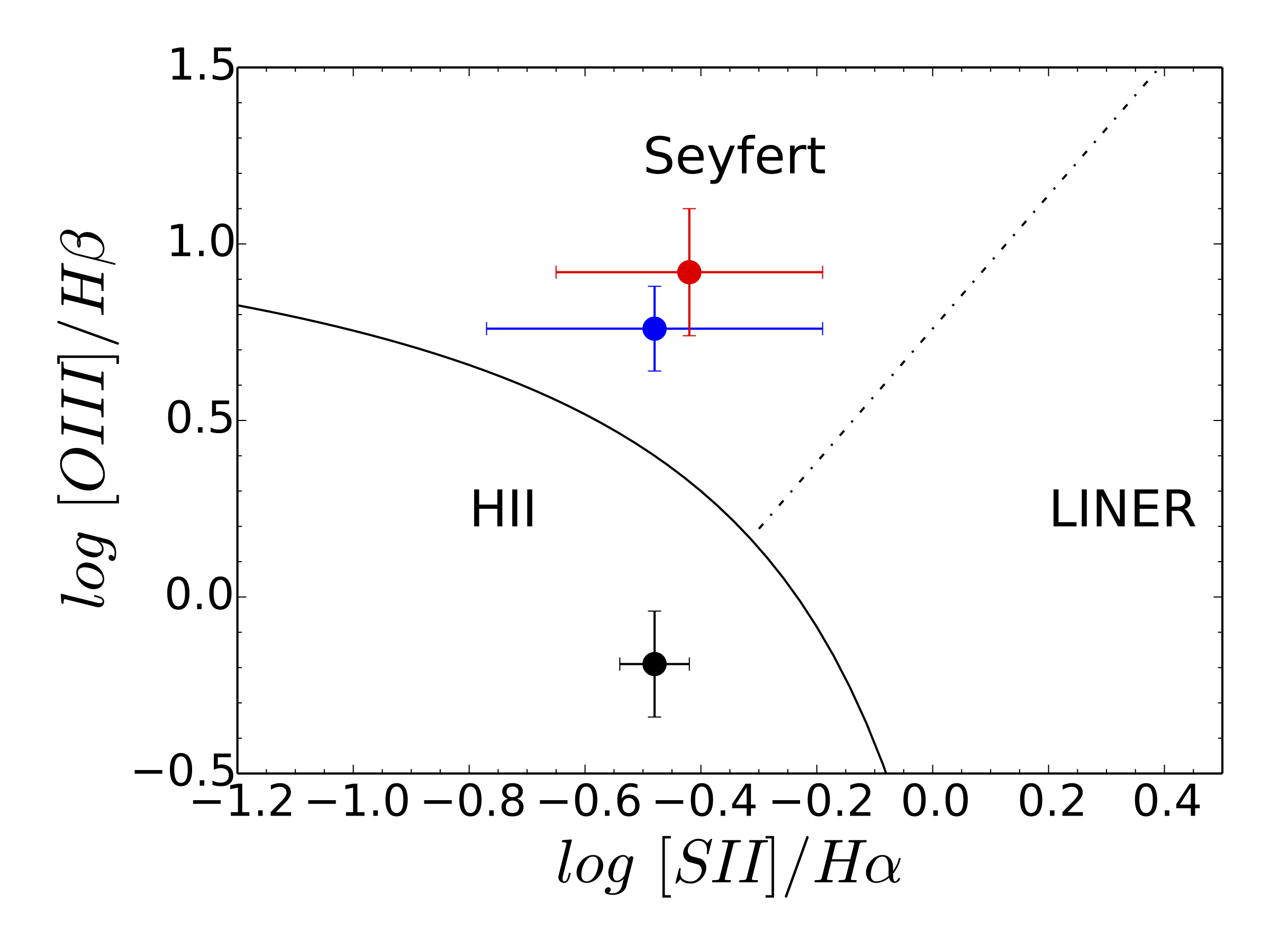}} 
\subfigure{\includegraphics[width=5.5cm,height=5cm]{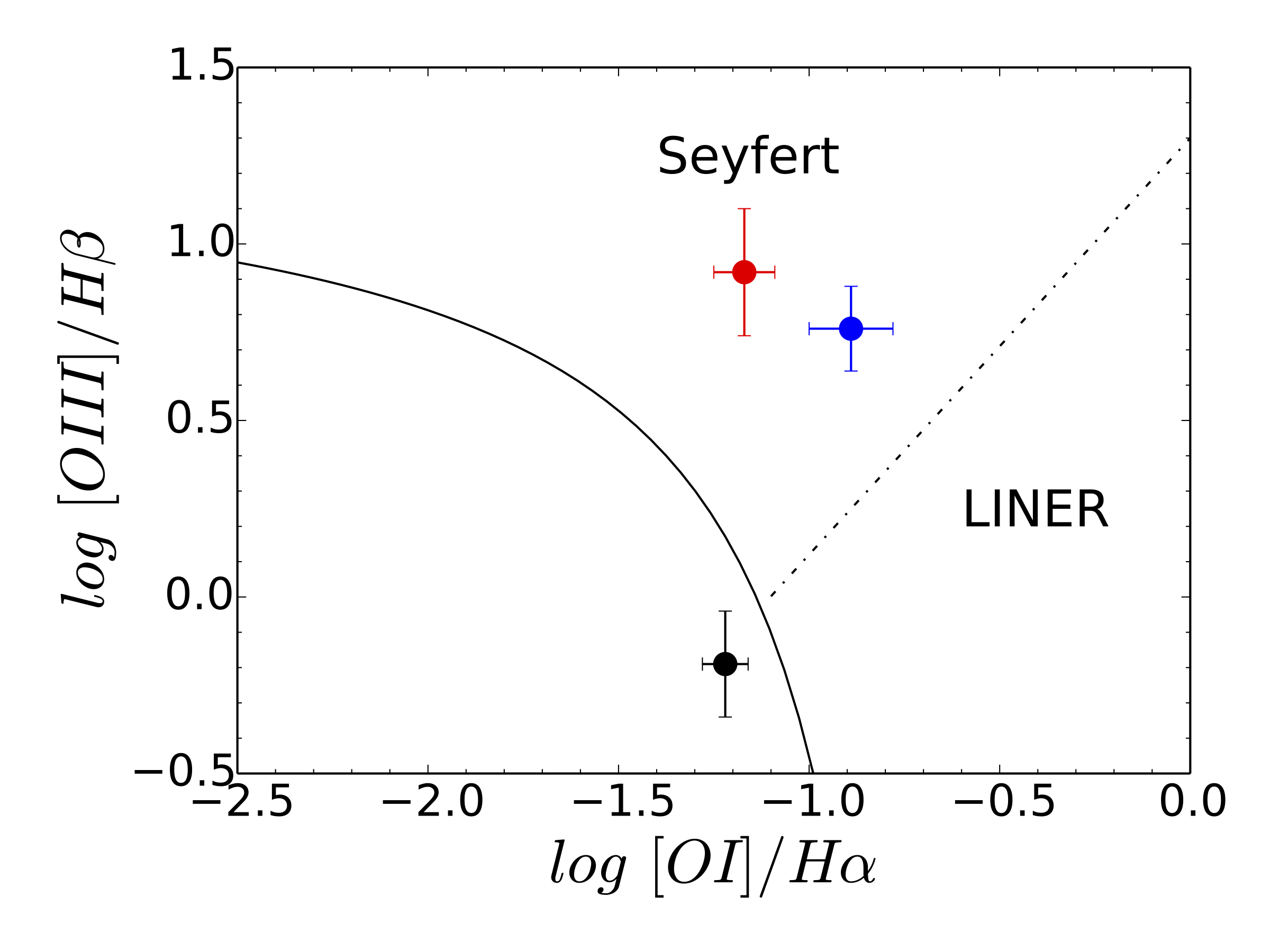}}
\caption{Spectral decomposition of the spectrum of Mrk\,622. Upper panels: \textit{WHT} profiles. The three panels show the best fit obtained with the 3G model. Black lines are the pure AGN spectrum obtained after using Starlight; blue and red lines show the blue and red shifted Gaussian components, the central Gaussian component is shown in black, while green lines show the best fit obtained. At the bottom of each panel the residuals are shown. Lower panels: \textit{BPT} diagrams. Solid black line marks the extreme Starburst (SB) classification proposed by \citet{2001ApJ...556..121K}, the dashed line marks the pure star formation region found by \citet{2003MNRAS.346.1055K}, and the dotted-dashed lines divide Seyfert/LINERS accordingly with \citet{2006MNRAS.372..961K}. Central, blue and red shifted components are shown in blue, red and black colours, respectively.
}
\label{3Gwht}
\end{figure*}


\subsection{Analysis of SDSS-DR7 spectrum}
\label{sdss}
In order to compare our results, a similar analysis was done with the \textit{SDSS}-DR7 \citep{2009ApJS..182..543A} public spectrum of Mrk\,622. The spectrum was obtained in MJD 52254 (see Table~\ref{obs}). The \textit{SDSS} pure AGN emission line spectrum was fitted following the same procedure as with the \textit{WHT} spectrum. For this spectrum the best fit was obtained using the 3G model. In the blue spectral region, the central Gaussian component resulted to be the narrowest of the three, with a FWHM of 314\,$\pm$\,31\,km\,s$^{-1}$. The other two blue and red shifted ones have FWHM\,=\,619\,$\pm$\,34\,km\,s$^{-1}$ and FWHM\,=\,523\,$\pm$\,48\,km\,s$^{-1}$, respectively. In the red spectral region, the central Gaussian component has a FWHM of 276\,$\pm$\,5\,km\,s$^{-1}$, and for the blue and red shifted components the estimated FWHM were 513\,$\pm$\,12\,km\,s$^{-1}$ and 548\,$\pm$\,31\,km\,s$^{-1}$, respectively.

\subsection{Optical classification}
\label{class}
Using the BPT diagnostic diagrams \citep[see][]{1981PASP...93....5B}, an optical empirical classification for Mrk\,622 was obtained. Therefore, three BPT diagrams were built using the line ratios reported in Table~\ref{ratios}, obtained with the \textit{WHT} modelled data. The loci of the intensity ratios obtained in these diagrams are in fairly good agreement. In all cases, the blue and red shifted narrow components appear in the AGN region, whereas the central narrow-line component appears in one diagram in the composite region, and in the other two in the SB region, see central panels of Figure~\ref{3Gwht}. The BPT diagrams obtained for the \textit{SDSS} data are not shown here because of their similarity in terms of both modelling and locus of the line ratios (c.f., Table~\ref{ratios}) in the BPT diagrams. The loci of the three components in all the BPT diagrams show that Mrk\,622 is a composite AGN (Sy2+SB) plus two AGN that are found to be spatially separated by $\sim$76 pc.

\section{Discussion and Conclusions}
\label{dis}
The optical long-slit spectroscopic observations herein presented, show that Mrk\,622 has three Gaussian components for the NLR, i.e. it is a triple-peaked object. Two out of the three Gaussian components, are separated by $\sim$500 km\,s$^{-1}$ and lie in the AGN loci of the BPT diagram, while the third one (central component) shows Composite AGN line ratios. The spatial separation of the two AGN components is only $\sim$\,76 pc. 

These results can be interpreted with different scenarios, namely: 1) A rotating disk or ring where double peaked lines are produced by a single ionizing source. In this case, \citet{2012ApJ...752...63S} suggest that equal flux double-peaked line-ratios are expected.  Since Mrk\,622 is Compton Thick, asymmetrical obscuration could be changing these ratios, so the validation of this scenario is not straightforward. 2) Since both SB galaxies and AGN can produce parsec scale winds/outflows, the blue and red components could be due to outflows. Following the proposed kinematic classification given by \citet{2016ApJ...832...67N}, and based in our results, Mrk\,622 could be an outflow composite source.  Note that in this second option the rotation scenario is discarded. In favour of this option is the large velocity offsets between the \oiiill \AA\ peaks. However, it must be noticed that the classical asymmetric profiles usually seen in outflows are not observed in any of the prominent emission lines. 3) A jet-driven outflow scenario is also plausible, since \citet{2003ApJS..148..327S} have found in an HST image that there is \oiiill \AA\ emission with an extent of 0.95$\times$1.3\arcsec at a PA of 55$^{\circ}$, perpendicular to the host galaxy major axis. To support this scenario, it would be required that the \oiii\ emission coincides with the radio jet orientation. New radio observations needed to check up this option are needed. 4) Finally, a scenario consisting of a binary AGN, with a spatial separation of $\sim$76\,pc, and a central AGN-starburst composite is also a possibility. 

To decide upon the various options allowed by the three peaks observed in the optical spectrum, high spatial resolution radio continuum observations of Mrk\,622 would be required. These will reveal whether there is a jet that follows the direction of the \oiiill \AA\ emission, which will favour/disfavour the jet-driven scenario. On the other hand, if the three sources have radio emission continuum and AGN characteristics, then this will support the presence of a binary AGN, surrounding a central AGN-Starburst composite.  The analysis of radio (\textit{VLA}), mid-Infrared (\textit{Canaricam}) and X-ray archival data of Mrk\,622 will be presented in a forthcoming paper (Ben\'itez et al., in preparation). 

\bigskip
We thank the anonymous referee for a critical reading of the paper and valuable suggestions. EB, ICG, JMRE, OGM, EJB, and CAN acknowledge support from DGAPA-UNAM grants IN111514 and IN113417. JMRE acknowledges support from the Spanish MINECO grant AYA2015-70498-C2-1-R. OGM thanks support from DGAPA-UNAM grant IA100516. CAN thanks support from DGAPA-UNAM grant IN107313 and CONACYT project 221398. DRD acknowledges support from the Brazilian funding agencies CNPq and CAPES. LG thanks support from CONACYT project 167236. EJB acknowledges support from grant IN109217. The \textit{WHT} is operated on the island of La Palma by the Isaac Newton Group in the Spanish Observatorio del Roque de los Muchachos of the Instituto de Astrof\'isica de Canarias. Our thanks to the supporting staff at WHT during the observing run. Funding for the Sloan Digital Sky Survey (SDSS) and SDSS-II has been provided by the Alfred P. Sloan Foundation and the Participating Institutions, the National Science Foundation, the U.S. Department of Energy, the National Aeronautics and Space Administration, the Japanese Monbukagakusho, and the Max Planck Society, and the Higher Education Funding Council for England. The SDSS Web site is \url{http://www.sdss.org/}. 

\bibliographystyle{mnras}
\bibliography{beniteze}

\bsp	
\label{lastpage}
\end{document}